\documentclass[aps,pra,amsmath,twocolumn,amssymb,showpacs]{revtex4}
\usepackage{txfonts}
\usepackage{epsfig,bm,dcolumn}
\usepackage{graphicx}
\usepackage{color}
\usepackage{overpic}

\begin{document}
\title{Medium Effects on Charmonium Production at LHC}
\author{Kai Zhou$^1$, Nu Xu$^{2,3}$, Zhe Xu$^1$, Pengfei Zhuang$^1$}
\address{$^1$Physics Department, Tsinghua University and Collaborative Innovation Center of Quantum Matter, Beijing 100084, China\\
         $^2$Nuclear Science Division, Lawrence Berkeley National Laboratory, Berkeley, CA 94720, USA\\
         $^3$Key Laboratory of Quark and Lepton Physics (MOE) and Institute of Particle Physics, Central China Normal University, Wuhan 430079, China}
\date{\today}
\begin{abstract}
We investigate with a transport approach the cold and hot nuclear matter effects on the charmonium
transverse momentum distributions in relativistic heavy ion collisions. The newly defined nuclear
modification factor $r_{AA}=\langle p_{T}^2\rangle_{AA}/\langle p_{T}^2\rangle_{pp}$
and elliptic flow $v_2$ for $J/\psi$ are sensitive to the nature of the
hot medium and the thermalization of heavy quarks. From SPS through RHIC to LHC colliding
energies, we observe dramatic changes in the centrality dependence of $r_{AA}$. We find that at LHC energy, the finally observed
charmonia are dominated by the regeneration from thermalized heavy quarks.
\end{abstract}
\maketitle

\section{Introduction}
\label{s1}
Charmonium production has long been considered as a clean probe for the
Quantum Chromodynamics (QCD) matter formed in heavy ion collisions
at relativistic energy~\cite{matsui}, due to the color screening of
heavy quark potential at finite temperature~\cite{lat1,lat2}. When the
temperature of the medium becomes higher than the charmonium dissociation temperature
$T_d$, the initially produced charmonia will be destroyed. It is called
the anomalous suppression~\cite{anomalous}. Therefore the comparison of the
final yield of charmonia from heavy ion collisions to that from the
corresponding nucleon-nucleon collisions can be used to extract the medium properties
in high-energy nuclear collisions~\cite{matsui}. This anomalous
suppression induced by the hot nuclear matter effect
explains well the experimentally observed
charmonium suppression at the Super Proton Synchrotron (SPS)
~\cite{na50,2na50}. However, color screening is not the only hot medium
effect in high-energy nuclear collisions. Once the energy is sufficiently
high, such as the collisions at the Relativistic Heavy Ion Collider (RHIC) and
the Large Hadron Collider (LHC), many charm quarks are produced~\cite{cstar,cphenix}.
The initially uncorrelated charm quarks $c$ and $\bar c$ can recombine into a
charmonium. This process is called regeneration~\cite{PBM, Rapp,Thews}. In high-energy
nuclear collisions, these two effects are co-existing and their relative strength
is energy dependent. At RHIC it turns out that both
suppression and regeneration processes are important for understanding
the charmonium production~\cite{yan,rapp1}.

Even before the formation of Quark-Gluon Plasma (QGP), the cold nuclear matter effects
affect the charmonium production. Usually three effects are considered
in the literatures:
(1) The change in the parton distribution function in nucleus which controls the initial
parton behavior and strongly depends on the collision kinematics. In
the small $x$ region, the nuclear parton distribution function
is clearly suppressed compared to that of nucleon. That is called the shadowing
effect~\cite{shadowing}; (2) Cronin effect~\cite{cronin} which describes the initial gluon
multi-scattering with the nucleons prior to any hard scattering and the quarkonium formation.
The nature of such collisions can be described as Brownian motion. As a result the transverse momentum distribution of the produced charmonia
is broadened; and (3) Nuclear absorption via interaction between charmonia and the primary nucleons which leads to the normal suppression of charmonia~\cite{absorp}. Both Cronin
effect and nuclear absorption have been studied in p+A collisions~\cite{pa,na50}, while
a deep understanding of shadowing effect requires e+p and e+A collisions. In order
to understand the charmonium production and extract properties of the medium
created in high-energy nuclear collisions, one must take into account
both cold and hot nuclear matter effects.

LHC provides new data on
the charmonium production in Pb+Pb collisions at colliding energy
$\sqrt {s_{NN}}=2.76$ TeV~\cite{cms1,alicemid,cms2,aliceraaptbin}.
At LHC the initially produced
charmonia are strongly suppressed by the hotter, larger and longer lived
fireball, hence the regeneration becomes the dominant production
source. Different from RHIC where the parton longitudinal momentum fraction $x\gtrsim 0.01$ is located in the interface between shadowing and anti-shadowing~\cite{lourenco} regions and therefore the shadowing and anti-shadowing effect
is not yet very strong (the ratio of parton distribution function in A+A and p+p collisions is between 0.96 and 1.08 for charmonium transverse momentum $0<p_t<5$ GeV/c, see Fig.7 of Ref.\cite{lourenco}), the
collisions at LHC with much smaller $x$ are in the strong shadowing region~\cite{eks98}, and
the shadowing effect plays an important role in the study of charmonium
production.

In the present study, we will extend our detailed transport approach for
charmonium motion in QGP to self-consistently including both the
cold and hot nuclear matter effects, and apply it to the charmonium
production at LHC energy. The suppression and regeneration
mechanisms in the hot medium are reviewed in Section \ref{s2}, and
the shadowing effect, the Gaussian smearing treatment for the Cronin
effect, and the initial charmonium distribution are focused in
Section \ref{s3}. The comparison between the theoretical
calculations and the experimental data are shown in Section
\ref{s4}. We summarize our study in Section \ref{s4}.

\section{Hot Nuclear Matter Effect}
\label{s2}
In order to extract information about the nature of the medium from
charmonium production in heavy ion collisions, the medium created in the initial stage and
the charmonia produced in the initial stage and in the medium should be treated both dynamically.

We employ the well tested 2+1 dimensional version~\cite{xiangzhuang} of the
ideal hydrodynamic equations
\begin{equation}
\label{hydro}
\partial_\mu T^{\mu\nu}=0
\end{equation}
to simulate the evolution of the almost baryon-free medium created
at RHIC and LHC, where $T_{\mu\nu}$ is the energy-momentum tensor of the
medium. The solution of the hydrodynamic equations provides the local
temperature and fluid velocity of the medium which will be used in the
charmonium regeneration and suppression.

To close the hydrodynamical equations one needs to know the equation
of state of the medium. We follow Ref.\cite{sollfrank} where the
deconfined phase at high temperature is an ideal gas of gluons and massless
$u$ and $d$ quarks plus 150 MeV massed $s$ quarks, and the
hadron phase at low temperature is an ideal gas of all known
hadrons and resonances with mass up to 2 GeV~\cite{pdg}. There is
a first order phase transition between these two phases. In the
mixed phase, the Maxwell construction is used. The mean field
repulsion parameter and the bag constant are chosen as $K$=450
MeV fm$^3$ and $B^{1/4}$=236 MeV to obtain the
critical temperature $T_c=165$ MeV~\cite{sollfrank} at vanishing baryon number
density.

Taking the initialization in Ref.~\cite{hirano}
for the hydrodynamic equations, we get the maximum temperature
$T_0=484$ and $430$ MeV of the medium at the initial time
$\tau_0=0.6$ fm/c, corresponding respectively to the observed charge number
density $dN_{ch}/dy=1600$ in mid rapidity and $1200$ in forward
rapidity by the ALICE collaboration~\cite{chargealice}.

Since a charmonium is so heavy, its equilibrium with the medium can
hardly be reached, one usually uses a transport approach to describe
the charmonium motion in the medium. The charmonium distribution
function $f_\Psi({\bf x},{\bf p},t|{\bf b})$ in the phase space
$({\bf x},{\bf p})$ at time $t$ in heavy ion collisions with impact
parameter ${\bf b}$ is
controlled by the Boltzmann-type equation
\begin{equation}
p^\mu \partial_\mu f_\Psi = - C_\Psi f_\Psi + D_\Psi .
\label{trans1}
\end{equation}
Considering the contribution from the feed-down of the excited
states $\psi'$ and $\chi_c$ to the finally observed
$J/\psi$~\cite{decay}, we should take into account the transport equations
for all the charmonium states $\Psi=J/\psi,\ \psi'$ and $\chi_c$, when we
calculate the $J/\psi$ distribution. The lose and gain terms $C_\Psi({\bf x},{\bf p},t|{\bf
b})$ and $D_\Psi({\bf x},{\bf p},t|{\bf b})$ describe the charmonium
dissociation and regeneration, and the elastic scattering is neglected since the charmonium mass is
much larger than the typical medium temperature. Introducing
proper time $\tau=\sqrt {t^2-z^2}$, space rapidity
$\eta=1/2\ln\left[(t+z)/(t-z)\right]$, momentum rapidity
$y=1/2\ln\left[(E+p_z)/(E-p_z)\right]$ and transverse energy
$E_T=\sqrt {E^2-p_z^2}$ to replace $t, z, p_z$ and $E=\sqrt{m^2+{\bf
p}^2}$, the transport equation can be rewritten as
\begin{equation}
\left[\cosh(y-\eta){\frac{\partial}{\partial\tau}}+{\frac{\sinh(y-\eta)}{\tau}}{\frac{\partial}{\partial
\eta}}+{\bf v}_T\cdot\nabla_T\right]f_\Psi
=- \alpha_\Psi f_\Psi+\beta_\Psi,
\label{trans2}
\end{equation}
where the third term in the square bracket arises from the free
streaming of $\Psi$ with transverse velocity ${\bf v}_T={\bf
p}_T/E_T$ which leads to a strong leakage effect at SPS
energy~\cite{spsleakage}, and the loss term $\alpha_\Psi({\bf x},{\bf
p},t|{\bf b}) = C_\Psi({\bf x},{\bf p},t|{\bf b})/E_T$ and gain term
$\beta_\Psi({\bf x},{\bf p},t|{\bf b}) = D_\Psi({\bf x},{\bf
p},t|{\bf b})/E_T$ are respectively the charmonium dissociation and
regeneration rates.

Considering the gluon dissociation $\Psi + g \to c+\bar c$ as the
dominant dissociation process in the hot QGP, $\alpha$ can be
calculated by the gluon momentum integration of the dissociation
cross section $\sigma_{g\Psi}$ multiplied by the thermal gluon
distribution $f_g$ and the flux factor $F_{g\Psi}$,
\begin{equation}
\label{loss}
\alpha_\Psi=\frac{1}{2E_T}\int
{d^3{\bf k}\over (2\pi)^3
2E_g}\sigma_{g\Psi}({\bf p},{\bf k},T)4F_{g\Psi}({\bf p},{\bf k})f_g({\bf k},T),
\end{equation}
where $E_g$ is the gluon energy, the cross section $\sigma_{g\Psi}$ in vacuum can be
derived through the operator production expansion with perturbative
Coulomb wave function~\cite{ope1,ope2,ope3,ope4}, and the cross section at
finite temperature is estimated by taking the geometrical relation
between the cross section and the average size of the charmonium
state, $\sigma_{g\Psi}({\bf p},{\bf k},T)=\sigma_{g\Psi}({\bf p},{\bf k},0) \langle
r_\Psi^2\rangle(T)/\langle r_\Psi^2\rangle(0)$. The averaged radius
square $\langle r_\Psi^2\rangle$ for the bound state $\Psi$ can be
obtained from the potential model~\cite{pot1,pot2}, its divergence
self-consistently defines a Mott dissociation temperature $T_d$
which indicates the melting of the bound state due to color
screening. The local temperature $T({\bf x},t|{\bf b})$ and the
fluid velocity $u_\mu({\bf x},t|{\bf b})$ of the medium in the cross section and the gluon thermal distribution come from the solution of the ideal hydrodynamic equations (\ref{hydro}).

The regeneration cross section is connected to the dissociation cross
section $\sigma_{g\Psi}$ via
the detailed balance between the gluon dissociation and
reversed process~\cite{yan}. To obtain the regeneration rate $\beta_\Psi$, we
need also the charm quark distribution $f_c({\bf x},{\bf q},t|{\bf
b})$ in the medium. From the experimental data at RHIC and LHC, the observed
large quench factor~\cite{rhicdquench,dquench} and elliptic flow~\cite{rhicdv2,dv2} for charmed
mesons indicate that the charm quarks interact strongly with the
medium. Therefore, one can reasonably take, as a good approximation, a
kinetically thermalized phase space distribution $f_c$ for charm quarks, $f_c({\bf x},{\bf q},t|{\bf b})=1/\left(e^{q^\mu u_\mu/T}+1\right)$.
Neglecting the creation and annihilation for charm-anticharm pairs
inside the medium, the spacial density of charm (anti-charm) quark
number  $\rho_c({\bf x},t|{\bf b})=\int d^3{\bf q}/(2\pi)^3f_c({\bf
x},{\bf q},t|{\bf b})$ satisfies the conservation law
\begin{equation}
\label{cflow}
\partial_\mu\left(\rho_c u^\mu\right)=0
\end{equation}
with the initial density determined by the nuclear geometry
\begin{equation}
\label{rhoc}
\rho_c({\bf x},\tau_0|{\bf b})=\frac{T_A({\bf x}_T)T_B({\bf x}_T-{\bf b})\cosh\eta}
{\tau_0} {d\sigma_{pp}^{c\bar c}\over d\eta},
\end{equation}
where $T_A$ and $T_B$ are the thickness functions at transverse
coordinate ${\bf x}_T$ defined in the Glauber model~\cite{glauber}, and
$d\sigma_{pp}^{c\bar c}/d\eta$ is the rapidity distribution of charm
quark production cross section in p+p collisions~\cite{alice4ccbar,fonll,fonll2,raplhc}.

\section{Cold Nuclear Matter Effect}
\label{s3}
The initial condition for the charmonium transport equation (\ref{trans2})
can be obtained from a geometrical superposition of
p+p collisions, along with the modification
from the cold nuclear matter effect. For
the charmonium production, the cold nuclear matter effect includes
usually the nuclear shadowing~\cite{shadowing}, Cronin effect~\cite{cronin} and nuclear absorption~\cite{absorp}. At LHC energy, the collision time for two heavy nuclei to pass through each other is
much shorter than the charmonium formation time and the QGP
formation time, and one can safely neglect the nuclear absorption
and just take into account the nuclear shadowing and Cronin effect.

The Cronin effect broadens the momentum of the initially produced
charmonia in heavy ion collisions. Before two gluons fuse into a
charmonium, they acquire additional transverse momentum via
multi-scattering with the around nucleons, and this extra momentum
would be inherited by the produced charmonium. Inspired from a
random-walk picture, we take a Gaussian smearing\cite{rapp1,liub} for
the modified transverse momentum distribution $\overline
f^{pp}_\Psi({\bf x},{\bf p},z_A,z_B|{\bf b})$
\begin{equation}
\label{cronin}
\overline f^{pp}_\Psi={1\over \pi a_{gN} l} \int
d^2{\bf p}_T' e^{-{\bf p}_T^{'2}\over a_{gN} l}f^{pp}_\Psi(|{\bf
p}_T-{\bf p}_T'|,p_z),
\end{equation}
where $l({\bf x},z_A,z_B|{\bf b})$ is the path length of the two gluons in
nuclei before their fusion into a charmonium at ${\bf x}$, $z_A$ and $z_B$ are the longitudinal coordinates of the two nucleons where the two gluons come from, $a_{gN}$ is the averaged charmonium transverse
momentum square obtained from the gluon scattering with a unit of length of nucleons, and $f^{pp}_\Psi({\bf p})$ is the
momentum distribution for a free p+p collision. The length
$l$ is calculated from the nuclear geometry, and the Cronin parameter
$a_{gN}$ is usually extracted from corresponding p+A collisions where the produced charmonia suffer from only
cold nuclear matter effect. Considering the absence of p+A data at LHC
energy, we take $a_{gN}=0.15$ GeV$^2$/fm from empirical estimations~\cite{emp,emp1,emp2}.

Assuming that the emitted gluon in the gluon fusion process $g+g\to
\Psi+g$ is soft in comparison with the initial gluons and the
produced charmonium and can be neglected in kinematics,
corresponding to the picture of color evaporation model at
leading order~\cite{Fritzsch:1977ay,cem1,cem2}, the longitudinal
momentum fractions of the two initial gluons are
calculated from the momentum conservation,
\begin{equation}
x_{1,2}={\sqrt{m_\Psi^2+p_T^2}\over \sqrt{s_{NN}}} e^{\pm y},
\end{equation}
where $y$ is the charmonium rapidity. The free distribution $f_\Psi^{pp}({\bf p})$ can
be obtained by integrating the elementary partonic processes,
\begin{equation}
\label{fg}
{d\sigma_\Psi^{pp}\over dp_Tdy}= \int dy_g x_1 x_2 f_g(x_1,\mu_F)
f_g(x_2,\mu_F) {d\sigma_{gg\to\Psi g}\over d\hat t},
\end{equation}
where $f_g(x,\mu_F)$ is the gluon distribution in a free proton, $y_g$ is the emitted
gluon's rapidity, $d\sigma_{gg\to\Psi g}/ d\hat t$ is
the charmonium momentum distribution produced from a fusion process, and
$\mu_F$ is the factorization scale of the fusion process.

Now we consider
the shadowing effect. The distribution function $\overline
f_i(x,\mu_F)$ for parton $i$ in a nucleus differs from a superposition of the
distribution $f_i(x,\mu_F)$ in a free nucleon. The nuclear shadowing
can be described by the modification factor $R_i=\overline f_i/(Af_i)$.
To account for the spatial dependence of the shadowing in a finite
nucleus, one assumes that the inhomogeneous shadowing is
proportional to the parton path length through the nucleus~\cite{shadpath},
which amounts to consider the coherent interaction of the incident
parton with all the target partons along its path length. Therefore,
we replace the homogeneous modification factor $R_i(x,\mu_F)$ by an
inhomogeneous one ${\cal R}_i(x,\mu_F,{\bf x})$~\cite{vogtshad}
\begin{equation}
{\cal R}_i=1+A\left(R_i-1\right)T_A({\bf x}_T)/T_{AB}(0)
\end{equation}
with the definition $T_{AB}({\bf b})=\int d^2{\bf x}_T T_A({\bf x}_T) T_B({\bf x}_T-{\bf b})$.
We employ in the following the EKS98 package~\cite{eks98} to evaluate the homogeneous
ratio $R_i$, and the factorization scale is taken as
$\mu_F=\sqrt{m_\Psi^2+p_T^2}$.

Replacing the free distribution $f_g$ in (\ref{fg}) by the modified
distribution $\overline f_g=Af_g{\cal R}_g$ and taking into account the Cronin
effect (\ref{cronin}), we finally get the initial charmonium distribution for the
transport equation (\ref{trans2}),
\begin{eqnarray}
f_\Psi({\bf x},{\bf p},\tau_0|{\bf b})&=&{(2\pi)^3\over E_T\tau_0}\int dz_Adz_B\rho_A({\bf x}_T,z_A)\rho_B({\bf x}_T,z_B)\nonumber\\
&\times&{\cal R}_g(x_1,\mu_F,{\bf x}_T){\cal R}_g(x_2,\mu_F,{\bf x}_T-{\bf b})\nonumber\\
&\times&\overline f_\Psi^{pp}({\bf x},{\bf p},z_A,z_B|{\bf b}),
\end{eqnarray}
where $\rho_A$ and $\rho_B$ are the nucleon distribution functions in the two colliding nuclei. Now the only thing left is the distribution $f_\Psi^{pp}$ in a free p+p collision which can be fixed by data or by some empirical estimations.

\begin{figure}[htb]
{\includegraphics[width=0.45\textwidth]{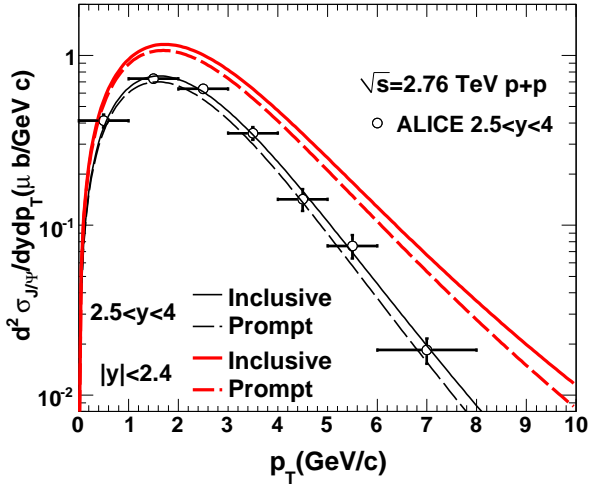}
\caption{(color online) The parametrization of inclusive (solid line) and prompt
(dashed line) $J/\psi$ production cross section as a function of transverse momentum
in 2.76 TeV p+p collisions. The thick-(red online) and thin-(black online) lines
represent mid rapidity and forward rapidity, respectively.
The data at forward rapidity are from the ALICE Collaboration
~\cite{alice2}.}
\label{fig1}
}
\end{figure}

Since the charmonia in heavy ion collisions at LHC are measured by
the CMS collaboration at mid rapidity $|y|<2.4$~\cite{cms1,cms2} and by
the ALICE collaboration at mid rapidity $|y|<0.9$~\cite{alicemid,alicemid2}
and forward rapidity $2.5<y<4$~\cite{alice1,alice12,alice13}, one should
consider the rapidity dependence of the free distribution $f_\Psi^{pp}$.
The charmonium states measured by the ALICE are reconstructed down to $p_T=0$ via the $\mu^+\mu^-$
decay channel at forward rapidity and via the $e^+e^-$ channel
at mid rapidity. From the measurement~\cite{alice2},
the averaged cross section is $d\sigma^{pp}_{\Psi}/dy=4.1\ \mu$b at $|y|<0.9$ and
$2.3\ \mu$b at $2.5<y<4$.  The ALICE has also measured the transverse
momentum distribution in p+p collisions~\cite{alice2}, the combined $y$ and $p_T$ dependence for inclusive $J/\psi$s at $2.5<y<4$ can be parameterized as
\begin{equation}
\label{ppjpsi}
{d^2\sigma^{pp}_\Psi\over dy p_Tdp_T}=\frac{2(n-1)}{(n-2)\langle p_T^2\rangle_{pp}}\left(1+{\frac{p_T^2}{(n-2)\langle p_T^2\rangle_{pp}}}\right)^{-n}{\frac{d\sigma^{pp}_\Psi}{dy}}
\end{equation}
with $n=5.06$ and $\langle p_T^2\rangle_{pp}=7.8$(GeV/c)$^2$, shown as the lower solid line in Fig.\ref{fig1}. From the ALICE~\cite{alice} and RHIC~\cite{rhic} data and CEM(Color Evaporation Model)~\cite{cem} calculation,
the free distribution $d^2\sigma^{pp}_\Psi/(dy p_Tdp_T)$ for inclusive $J/\psi$s at rapidity $|y|<2.4$
can also be described by the parametrization (\ref{ppjpsi}) with $n=4$,
$\langle p_T^2\rangle_{pp}=10$ (GeV/c)$^2$ and $d\sigma_\Psi^{pp}/dy=4.1\ \mu$b, see the other solid line in Fig.\ref{fig1}.
\begin{figure}[htb]
{\includegraphics[width=0.42\textwidth]{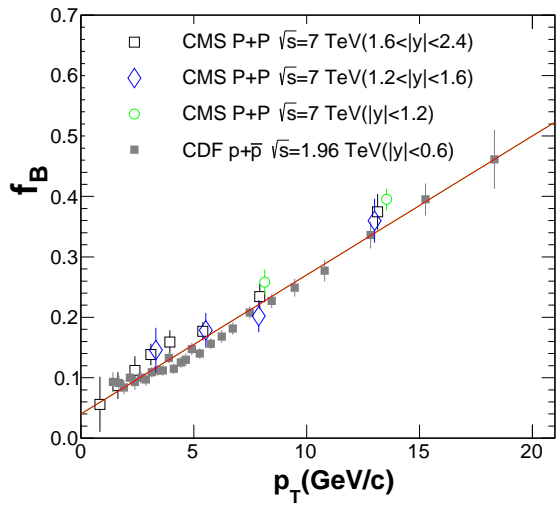}
\caption{(color online) The B decay fraction as a function of $J/\psi$
transverse momentum in p+p collisions. The data are from the
CDF~\cite{fbcdf} and CMS~\cite{fbcms} collaborations, and the straight line
is our linear parametrization. }
\label{fig2}}
\end{figure}

The inclusive $J/\psi$s measured by ALICE include the prompt part and
the contribution from the B decay. The former consists of direct production
and feed down from the excited states, $30\%$ from $\chi_c$ and
$10\%$ from $\psi'$, and the latter comes from the decay of bottomed hadrons. The B decay contributes about
$10\%$~\cite{bdy10,fbcdf,fbcms} to
the total inclusive yield. Since the ALICE does not separate the two parts
from each other, we have to take
into count the B decay contribution in our model to compare our theoretical calculation
with the ALICE data. CMS experiment has measured the prompt charmonia at high transverse momentum
$6.5<p_T<30$ GeV/c and at mid rapidity $|y|<2.4$~\cite{cms1}.
Since at the moment there are no p+p data for prompt charmonia in the CMS
rapidity region, we employ $d\sigma^{pp}_\Psi/dy=4.1\ \mu$b from the ALICE
data~\cite{alice2} for inclusive $J/\psi$s in mid rapidity, and then eliminate
the B decay contribution by multiplying the inclusive cross section by a factor of
$\left(1-f_B(p_T)\right)$, where $f_B$ is the B decay fraction. Fig.\ref{fig2}
shows the recent data
on the B decay fraction in p+p($\bar p$) collisions as a function of
$J/\Psi$ $p_T$~\cite{fbcms}. The data can be well parameterized as
$f_B(p_T)=0.04+0.023 p_T$/(GeV/c). Note that the linear parametrization
is rapidity independent in the region we considered.
The prompt $J/\psi$ distribution at mid rapidity $|y|<2.4$ is
shown as the upper dashed curve in Fig.\ref{fig1}. Since the B decay fraction is approximately rapidity independent,
the distribution $f_\Psi^{pp}$ for inclusive $J/\psi$s at mid rapidity $|y|<0.9$
can be obtained from the prompt distribution at mid rapidity $|y|<2.4$ and the B decay fraction $f_B$ shown in Fig.\ref{fig2}.

We now turn to the rapidity dependence of the charm quark production
cross section $\sigma_{c\bar c}^{pp}$ in (\ref{rhoc}). From the
ALICE data~\cite{alice4ccbar}, there is $d\sigma_{c\bar
c}^{pp}/dy=0.65$ mb at mid rapidity. By taking the
FONLL~\cite{fonll,fonll2} scaling, we extract $d\sigma_{c\bar
c}^{pp}/dy=0.4$ mb at forward rapidity. Considering the uncertainty in the experimental data and theoretical calculations (see Fig.5 of Ref.\cite{alice4ccbar}), the upper limits of  FONLL~\cite{fonll2} and
pQCD are often used in
models~\cite{raplhc}, to estimate the maximum quarkonium production. We take in the following numerical
calculations $d\sigma_{c\bar
c}^{pp}/dy$ between $0.65$ and $0.8$ mb at mid rapidity and $0.4$ and $0.5$ mb at forward
rapidity. Note that the shadowing effect changes not only the initial
$J/\psi$ distribution, but also the in-medium $J/\psi$ regeneration
and the non-prompt contribution from the B decay, by reducing the number
of charm and bottom quarks. In principle, the shadowing should be centrality dependent. To simplify the numerical calculations, we take in the following a reduction of $20\%$ for the charm and bottom quark production cross sections, estimated from the centrality averaged EKS98 evolution~\cite{eks98}.

\section{Numerical Results}
\label{s4}
\subsection{Centrality Dependence of $R_{AA}$}
\begin{figure}[htb]
\vspace{-0.2cm}
{\includegraphics[width=0.50\textwidth]{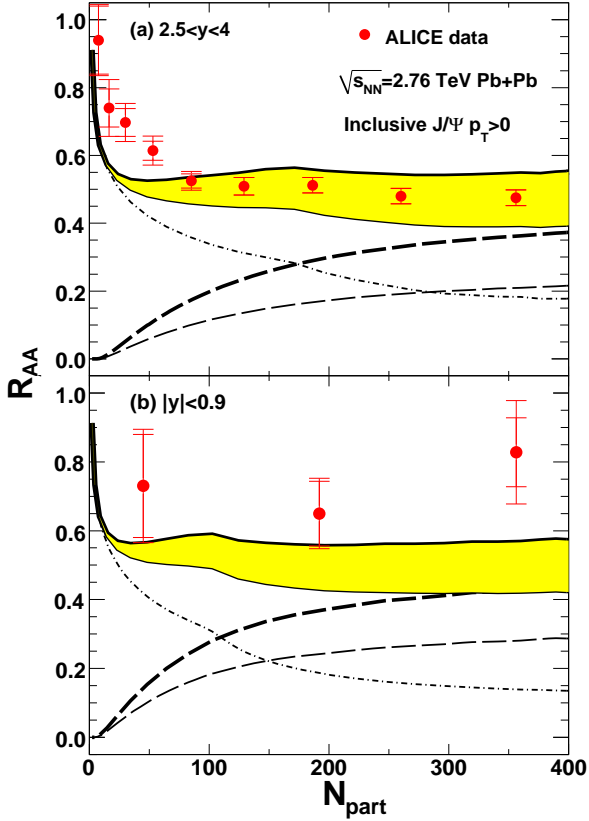}
\caption{(color online) The $J/\psi$ nuclear modification factor as a
function of centrality for 2.76 TeV Pb+Pb collisions at forward
rapidity (upper panel) and mid rapidity (lower panel). The dot-dashed lines
represent the initial fraction. The thick and thin dashed lines
represent the regeneration fraction with the upper and lower
limits of charm quark cross-sections $d\sigma_{c\bar c}^{pp}/dy=0.8$ and $0.65$ mb
at mid rapidity and $0.5$ and $0.4$ mb at forward rapidity.
The hatched bands represent the full results. The data points are taken from ALICE experiment~\cite{alice2}. }
\label{fig3}}
\end{figure}

We start with the $J/\psi$ nuclear modification factor $R_{AA}=N_{AA}/\left(N_{coll} N_{pp}\right)$ as a function of the number of participants $N_{part}$, where $N_{pp}$ and $N_{AA}$ are respectively the numbers of measured $J/\psi$s in p+p and A+A collisions, and $N_{coll}$ is the number of nucleon-nucleon collisions at fixed $N_{part}$. Our model calculations for 2.76 TeV Pb+Pb collisions and the comparison with the inclusive ALICE data are shown in Fig.\ref{fig3} at forward rapidity (upper panel) and mid rapidity (lower panel).
\begin{figure}[htb]
{\includegraphics[width=0.50\textwidth]{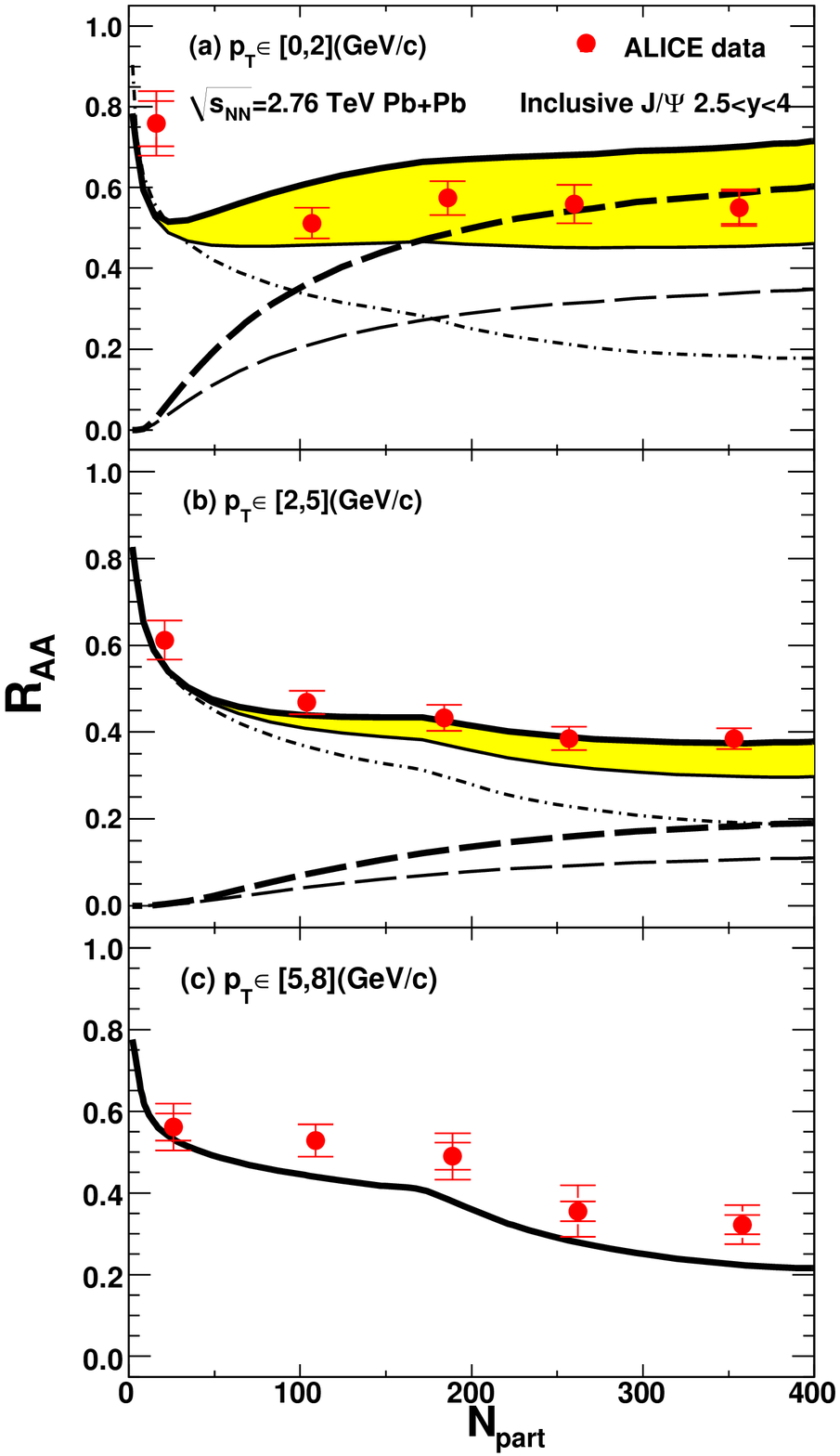}
\caption{(color online) The $J/\psi$ nuclear modification
factor as a function of centrality for 2.76 TeV Pb+Pb collisions
at forward rapidity and in different transverse momentum bins.
The dot dashed lines are the initial fraction, the thick and thin
dashed lines are the regeneration fraction with charm quark cross section $d\sigma_{c\bar c}^{pp}/dy=0.5$ and $0.4$ mb, and the bands are
the full result. The data are from the ALICE collaboration~\cite{aliceraaptbin}.}
\label{fig4}}
\end{figure}
\begin{figure}[htb]
{\includegraphics[width=0.45\textwidth]{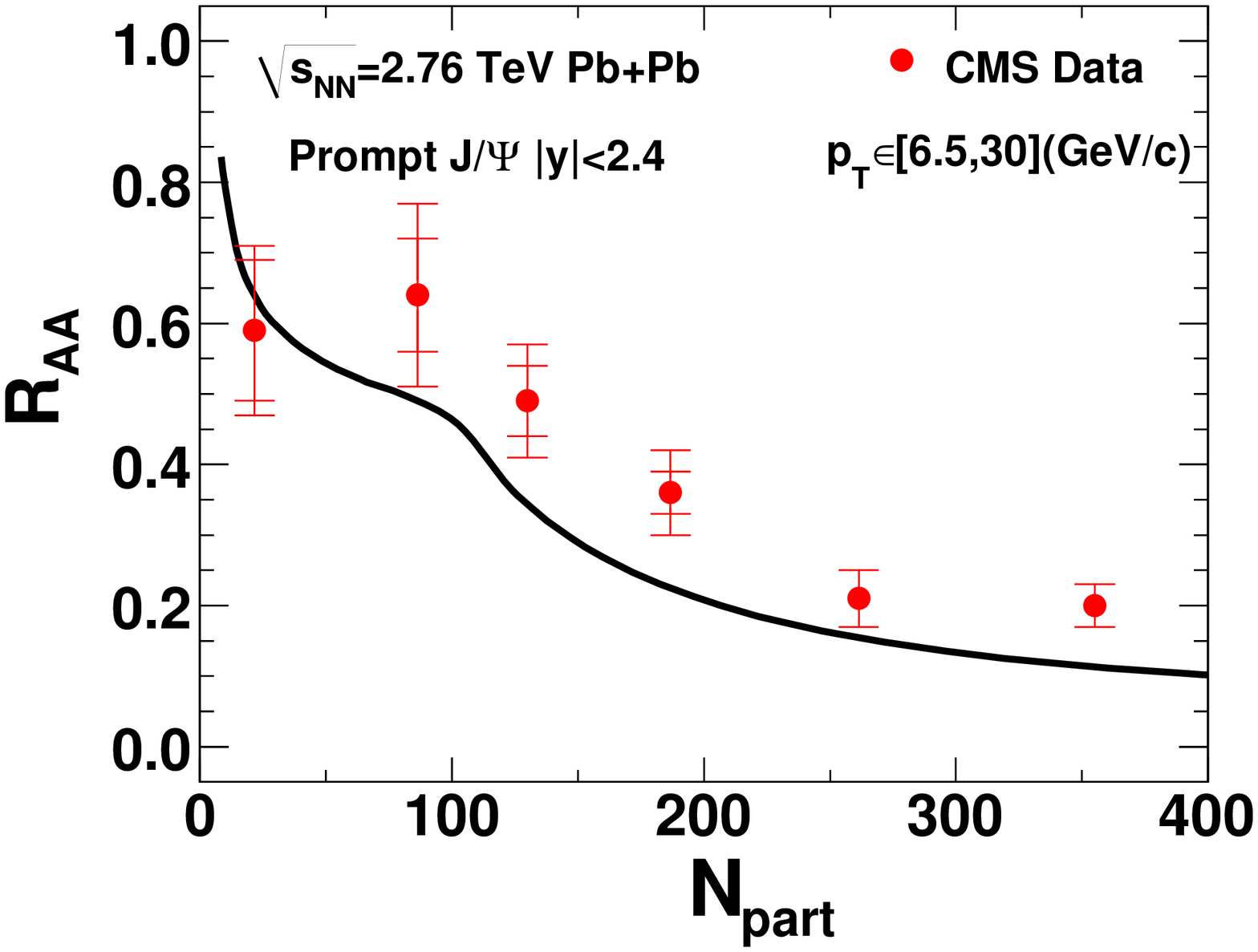}
\caption{(color online) The $J/\psi$ nuclear modification factor
as a function of centrality for 2.76 TeV Pb+Pb collisions at mid rapidity
and in a high transverse momentum bin. The line is the model calculation
and the data are from the CMS collaboration~\cite{cms1}.}
\label{fig5}}
\end{figure}

With increasing centrality, the initial contribution (dot-dashed lines) drops
down and the regeneration (dashed lines) goes up monotonously. Considering the
uncertainty in the charm quark production cross section, we take the range of cross-sections
$d\sigma^{pp}_{c\bar c}/dy=0.4$ and $0.5$ mb at forward rapidity and
$0.65$ and $0.8$ mb at midrapidity, corresponding to the thin and thick
dashed lines in Fig.\ref{fig3}. This uncertainty in the regeneration results
in a band for the full result. Different from the collisions at SPS energy
where the regeneration can be neglected~\cite{Rapp,xiangzhuang} and at RHIC energy where
the initial production is still a dominant component and the regeneration
becomes equivalently important only in very central collisions~\cite{yan,liu1},
the regeneration at LHC energy becomes the dominant source of charmonium
production in a wide centrality bin. The competition between the strong
dissociation and regeneration leads to a flat structure
for the total charmonium production at both forward rapidity and mid rapidity,
see Fig.\ref{fig3}. At forward rapidity where high statistic data are available,
our model results well explain the data in semi-central and
central collisions with $N_{part}\gtrsim 100$, and the deviation from the
data in small $N_{part}$ region is probably due to the invalidation of
hydrodynamics for the medium and the canonical limit for the charmonium regeneration~\cite{ralf,cononical}.
At mid rapidity where the data are with large error bars, the model results seem
systematically under the data.

We have seen the strong competition between the charmonium suppression and
regeneration in the centrality dependence of the $J/\psi$ yield. In order to see
the charmonium production and suppression mechanisms more clearly in heavy ion
collisions, we turn to the transverse momentum dependence of the yield.
Figs.\ref{fig4} and \ref{fig5} show the nuclear modification factor $R_{AA}$ as
a function of centrality in different $p_T$ bins. In addition we make comparisons
with ALICE data ~\cite{aliceraaptbin} (Fig.4) and CMS data ~\cite{cms1} (Fig.5) for
inclusive and prompt $J/\psi$ production, respectively.

In our calculation, we considered a reduction of $20\%$ for the charm quark number
due to the strong shadowing effect at LHC energy. We also took a kinetically thermalized charm quark
distribution, by assuming strong interaction between charm quarks and the medium.
For bottom quarks, we took into account the same reduction for the number in the
calculation of $p_T$ integrated yield.
Since bottom quarks are so heavy, their thermalization is unreasonable in nuclear collisions. While the total $J/\psi$ yield shown in Fig.\ref{fig3} is not sensitive to the bottom quark transverse momentum distribution, it may change remarkably the $J/\psi$ yield in a fixed $p_T$ bin, especially in a high $p_T$ bin
where the bottom quark energy loss becomes important. Inspired from the CMS measurement~\cite{cms2},
we apply a B quench factor $R_{AA}^B=0.4$ when calculating the non-prompt $J/\psi$s in the region
of $5<p_T<8$ GeV/c.

Heavy quarks are produced via hard scatterings and their initial momentum distribution
is hard. Through interation with the medium, heavy quarks lose energy, and the corresponding
$p_{t}$ distribution becomes steeper. Since the medium is hot and dense, some of the charm
quarks are even thermalized. Considering that the regenerated charmonia from thermalized charm
quarks in hot medium are mainly distributed in low momentum region, their contribution
to the yield decreases with increasing transverse momentum, and therefore the band
structure for the total result due to the uncertainty in regeneration disappears in
the high $p_T$ bins $5<p_T<8$ GeV/c and $6.5<p_T<30$ GeV/c, see Figs.\ref{fig4} and
\ref{fig5}. As we have pointed out that the flat platform for semi-central and central
collisions comes from the competition between the suppression and regeneration, it
vanishes in the high $p_T$ bins due to the disappearance of the regeneration. The
yield in high $p_T$ bins is characterized by the Debye screening effect on the
initially produced charmonia. In this case, there exists a kink for the nuclear
modification factor which is located at $N_{part}\sim 200$ (Fig.4) for $5<p_T<8$ GeV/c
at forward rapidity and $N_{part}\sim 100$ (Fig.5) for $6.5<p_T<30$ GeV/c in mid rapidity.
Before the kink, the temperature of the fireball is less than the $J/\psi$
dissociation temperature $T_{J/\psi}$ and there is $R_{AA}\sim 0.6$, resulted
from the fact that the decay contribution from the excited charmonium states to
the finally observed $J/\psi$s is about $40\%$. Starting at the kink, the temperature
is higher than $T_{J/\psi}$ and the $J/\psi$ suppression becomes more and more
important in more central collisions. As one can see from Figs.\ref{fig4} and \ref{fig5}, the kink
at mid rapidity appears at a lower value of $N_{part}$ than that at forward-rapidity.
This is due to the fact that the medium at mid rapidity is hotter and $J/\psi$ starts to
melt at more peripheral collisions. The model results underestimate the high $p_{T}$ results
a bit, see Figs.\ref{fig4} and \ref{fig5}. Part of the discrepancy may due to the fact we
have not include the velocity dependence of the dissociation temperature. As discussed
in~\cite{yunpengTdv}, fast moving charmonia tend to be melt at a higher temperature making
the velocity dependence particularly important in high $p_{T}$ regions.

\subsection{Transverse Momentum Dependence of $R_{AA}$}
\begin{figure}[htb]
{\includegraphics[width=0.50\textwidth]{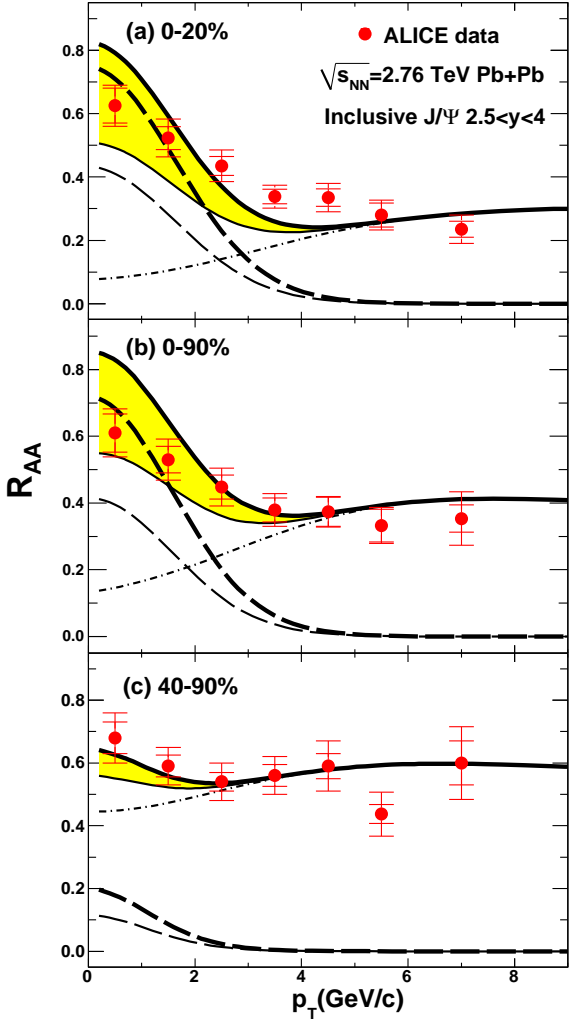}
\caption{(color online) The $J/\psi$ nuclear modification factor as a function
of transverse momentum for 2.76 TeV Pb+Pb collisions at forward rapidity and in
different centrality bins. The dot-dashed lines are the initial fraction, the thick
and thin dashed lines are the regeneration fraction with charm quark cross section $d\sigma_{c\bar c}^{pp}/dy=0.5$ and $0.4$ mb, and the bands are the full result. The data
are from the ALICE collaboration~\cite{aliceraaptbin}.}
\label{fig6}}
\end{figure}
\begin{figure}[htb]
{\includegraphics[width=0.45\textwidth]{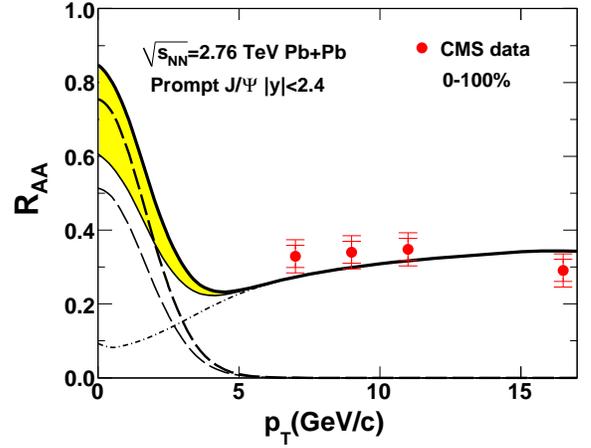}
\caption{(color online) The $J/\psi$ nuclear modification factor as a function
of transverse momentum for 2.76 TeV Pb+Pb collisions at mid rapidity and in minimum
bias event. The dot-dashed line is the initial fraction, the thick and thin dashed
lines are the regeneration fraction with charm quark
cross section $d\sigma_{c\bar c}^{pp}/dy=0.8$ and $0.65$ mb, and the band is the full result. The data are from the CMS
collaboration~\cite{cms2}.}
\label{fig7}}
\end{figure}
The transverse momentum distribution of the nuclear modification factor $R_{AA}(p_T)$ in a
fixed centrality bin and its comparison with the ALICE data are shown in Fig.\ref{fig6}
for inclusive $J/\psi$s at forward rapidity. We see clearly that, the regeneration (dashed
lines) dominants the low $p_{T}$ regions at all centralities. On the other hands, initially
produced $J/\psi$s (dot-dashed lines) become important at high $p_{T}$ region. Therefore, the
competition between the suppression and regeneration depends strongly on the transverse
momentum and collision centralities. When both the regeneration at low $p_T$ and initial production at high
$p_T$ are important, there exists a minimum structure located at intermediate $p_T$, see
the upper and middle panels of Fig.\ref{fig6}. Note that without the regeneration as the
second production source, the nuclear modification factor $R_{AA}$ would decrease monotonously
with increasing centrality and increases monotonously with increasing transverse momentum.
The $R_{AA}(p_T)$ for prompt $J/\psi$s at mid rapidity is shown in Fig.\ref{fig7}. While
the CMS data are in the high $p_T$ region where there is almost no contribution from the
regeneration and the system is controlled only by the dissociation of the initially
produced $J/\psi$s, the predicted $R_{AA}$ at very low $p_t$ is
larger than unity which is not possible without the regeneration mechanism.

\subsection{A New Ratio $r_{AA}=\langle p_T^2\rangle_{AA}/\langle p_T^2\rangle_{pp}$}
\begin{figure}[htb]
{\includegraphics[width=0.50\textwidth]{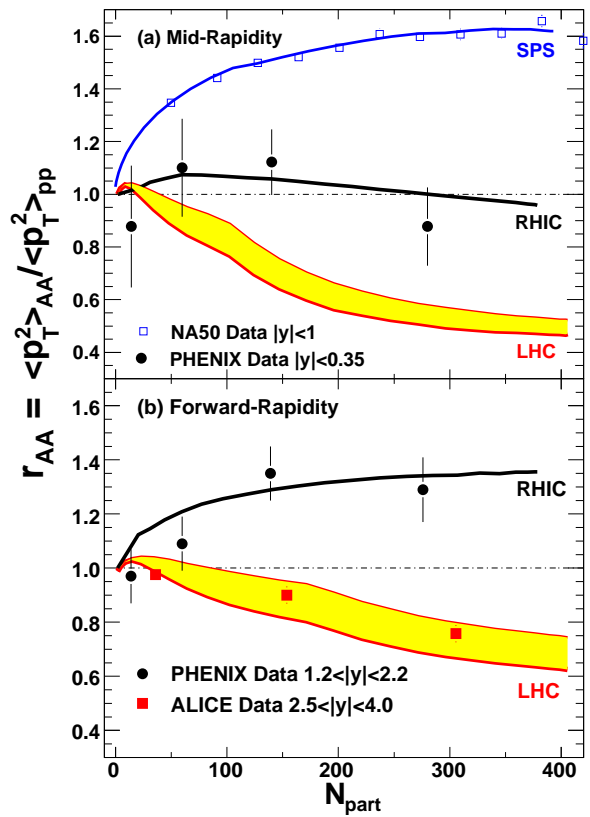}
\caption{(color online) The newly defined nuclear modification factor
$r_{AA}=\langle p^2_T \rangle _{AA}/\langle p^2_T\rangle _{pp}$ for $J/\psi$s
as a function of centrality at SPS, RHIC and LHC energies. The bands at LHC are due
to the uncertainty in the charm quark cross section $0.4<d\sigma_{c\bar c}^{pp}/dy<0.5$ mb at forward rapidity (lower panel) and $0.65<d\sigma_{c\bar c}^{pp}/dy<0.8$ mb at mid rapidity (upper panel), and the
data are from NA50~\cite{na50}, PHENIX~\cite{phenix} and ALICE~\cite{ptraaalice} collaborations. }
\label{fig8}}
\end{figure}

For heavy ion collisions in the past decades, the colliding energy from SPS
to LHC increases by two orders of magnitude, and the heavy quark production cross section
increases dramatically. While the transverse momentum dependence of the nuclear
modification factor $R_{AA}$ tells us the importance of the regeneration mechanism,
we hope to find a quantity which is more sensitive to the nature of the medium.
On the other hand, in our treatment for hot nuclear matter effect, we assumed a
thermalized charm quark distribution for the calculation of the regeneration rate,
how good is this assumption and how does the charm quark thermalization affect the
charmonium distribution? The other question is how to separate the hot nuclear
matter effect from the cold nuclear matter effect. To focus on the property of
the hot medium, we hope to have a quantity which is sensitive to the hot nuclear
matter effect but affected weekly by the cold nuclear matter effect. To these ends,
we introduced a new nuclear modification factor for the transverse momentum~\cite{zhou}
\begin{equation}
\label{raa1}
r_{AA}={\langle p_T^2\rangle_{AA}\over \langle p_T^2\rangle_{pp}},
\end{equation}
where $\langle p^2_t\rangle _{AA}$ and $\langle p^2_t\rangle _{pp}$
are averaged transverse momentum square for $J/\psi$ in A+A and p+p collisions, respectively.
To be different from the usually used nuclear modification factor $R_{AA}$ for the yield,
we take here $r_{AA}$ as the ratio of the second moment of transverse momentum distributions.
Since the nuclear matter effect changes not only the size of the transverse momentum but
also its shape, we focus on $\langle p_T^2\rangle$ instead of $\langle p_T\rangle$. It is
worthy to note that the second moment of the $p_{T}$ distribution is also sensitive to the
Cronin effect but not much affected by the parton distribution functions. The calculation
and the comparison with experimental data at SPS, RHIC and LHC energies are shown in Fig.\ref{fig8} for two rapidity bins. In mid rapidity where the hot nuclear matter effect is most important, the ratio changes significantly with increasing colliding energy,
\begin{equation}
\label{raa2}
r_{AA}\left\{ \begin{array}{ll}
>1 &\textrm{SPS}\\
\sim 1 &\textrm{RHIC}\\
<1 &\textrm{LHC}
\end{array} \right . \ \ \ \ \text {at mid rapidity},
\end{equation}
see the upper panel of Fig.\ref{fig8}. At SPS, almost all the measured $J/\psi$s are produced through
initial hard processes and carry high momentum. The continuous
increasing with centrality arises from the Cronin effect and
leakage effect~\cite{xiangzhuang}. The latter is described by the third term
on the left hand side of the transport equation (\ref{trans2}) and means that the high $p_T$ $J/\psi$s can escape from the hot medium. At RHIC, the regeneration starts
to play a role and even becomes equally important as the initial
production in central collisions~\cite{yan,liu1}, the cancelation between
the suppression and regeneration leads to a flat $r_{AA}$ around unity.
At LHC, the regeneration becomes dominant, especially in central collisions.
Since the regenerated charmonia carry low momentum in comparison with the initial
production, the more and more important regeneration results in a decreasing $r_{AA}$
with increasing centrality. This tremendous change in the $p_T$ ratio comes from the nature
of the hot medium at each corresponding collision energy. Our calculation agrees well
with the data at SPS and RHIC energies. At LHC energy there are currently no data at
mid rapidity. The band structure at LHC is again from the uncertainty in the charm quark
production cross section. Note that, different from the ratio $R_{AA}$ for the yield
where the lower and higher borders of the band correspond to the smaller and larger
cross sections, the relation between the two borders of $r_{AA}$ and the cross section
is in an opposite way, namely, the lower and higher borders here correspond to the larger
and smaller cross sections. Intuitively this is simple to understand: In the case of
larger charm quark cross section, more $J/\psi$s are from the regeneration and then carry low momentum. To have a comparison with the LHC data, we calculated the
ratio $r_{AA}$ in the forward rapidity where the hot medium effect becomes weaker.
In this case the trend of the ratio is still very different at RHIC and LHC: With
increasing centrality it goes up at RHIC but drops down at LHC.
\begin{equation}
\label{raa3}
r_{AA}\left\{ \begin{array}{ll}
>1 &\textrm{RHIC}\\
<1 &\textrm{LHC}
\end{array} \right .\ \ \ \ \text{at forward rapidity},
\end{equation}
see the lower panel of Fig.\ref{fig8}. The calculation agrees well with both the
RHIC and LHC data.
\begin{figure}[htb]
{\includegraphics[width=0.50\textwidth]{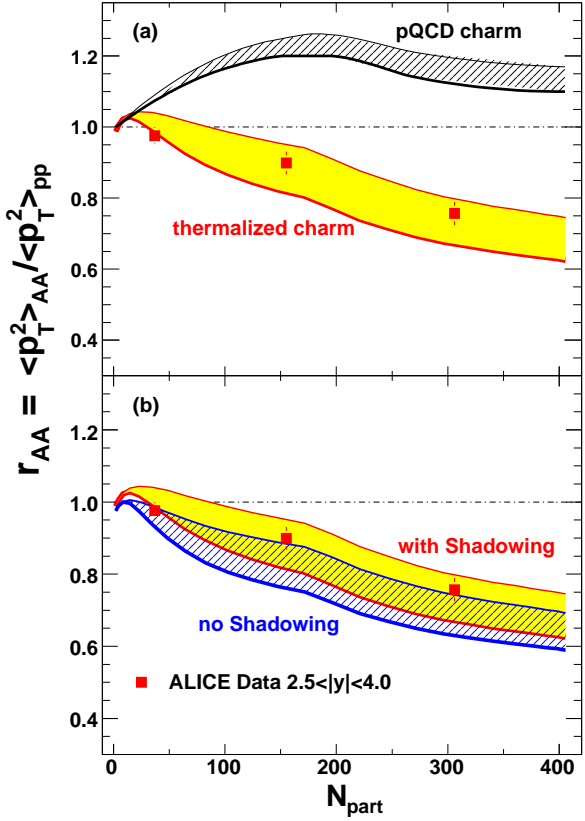}
\caption{(color online) The newly defined nuclear modification factor
$r_{AA}=\langle p^2_T \rangle _{AA}/\langle p^2_T\rangle _{pp}$
for $J/\psi$s as a function
of centrality at LHC energy and at forward rapidity. The upper and lower panels
are the comparisons between pQCD and thermal charm quark distributions and
between with and without shadowing
effect. The bands are again due to the uncertainty in the charm quark
cross section $0.4<d\sigma_{c\bar c}^{pp}/dy<0.5$ mb, and the data are from the ALICE collaboration~\cite{ptraaalice}.}
\label{fig9}}
\end{figure}

After a careful inspecting the centrality dependence of $r_{AA}$, one can find a common
feature in all cases, namely, $r_{AA}$ in peripheral collisions is always increasing as
the centrality increases. In case the initial production is dominant, the ratio continues
to increase, while strong regeneration from thermalized charm quarks pulls the ratio
down below unity in more central collisions. In order to see if charm quarks are thermalized
with the hot medium, we calculate now the ratio $r_{AA}$ at LHC energy with a pQCD simulated
charm quark distribution, shown in the upper panel of Fig.\ref{fig9} at forward rapidity.
The thermal distribution is the limit of strong interaction between charm quarks and the
medium, while the pQCD distribution taken from the simulator PYTHIA~\cite{pythia} is the
limit of no interaction. For the pQCD distribution,
the charm quark energy loss is excluded, the initially produced charm quarks keep their high
momentum in the medium, and therefore the regenerated charmonia will be no longer soft but
carry high momentum. In this case, the averaged transverse momentum is enhanced at the forward
rapidity and the ratio $r_{AA}$ becomes larger than the one with thermal distribution,
see the upper panel of Fig.\ref{fig9}. Again the band is due to the uncertainty in the charm
quark cross section. From the comparison with the ALICE data, we conclude that charm quarks
are thermalized in high-energy nuclear collisions at the LHC energy.

In the above calculations we have taken a reduction of $20\%$ for the charm and bottom quark distributions in the medium, due to the shadowing effect. This reduction leads to an extra charmonium suppression. Supposing the total number of initially produced charm quarks is $n_c$, the regenerated charmonium number is then proportional to $n_c^2$ if we do not take into account the shadowing effect and $(0.8n_c)^2=0.64n_c^2$ when the shadowing effect is included. However, the averaged transverse momentum square $\langle p_T^2\rangle$ is a normalized quantity, it should not be so sensitive to the shadowing effect. The comparison between the calculations with and without considering the shadowing effect is shown in the lower panel of Fig.\ref{fig9}. As we expected, the two bands are very close to each other and overlap partly. This indicates that the nuclear modification factor $r_{AA}$ for charmonium transverse momentum can be used to extract the hot medium information with only small correction from the early
shadowing effect.

\subsection{Elliptic Flow $v_2$}
\begin{figure}[htb]
{\includegraphics[width=0.50\textwidth]{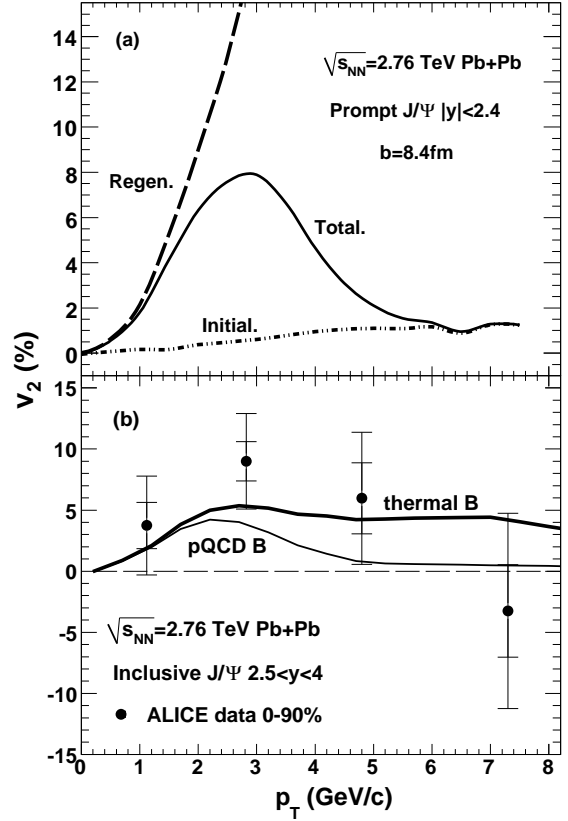}
\caption{(upper panel) The elliptic flow $v_{2}$ for prompt (upper panel) and inclusive (lower panel) $J/\psi$s
in $\sqrt {s_{NN}}=2.76$ TeV Pb+Pb collisions. The calculation is with
impact parameter $b=8.4$ fm, corresponding to the minimum bias event. The dot-dashed, dashed and solid lines represent the initial, regeneration and total contributions.
In the lower panel, the thick and thin lines indicate the total results including
thermal bottom decay and pQCD bottom decay, respectively. The experimental data are
taken from ALICE experiment~\cite{alicev2}.}
\label{fig10}}
\end{figure}
Since the elliptic flow originates from the geometric anisotropy in the configuration
space and develops in the evolution of the
hot medium, it is significant in semi-central collisions and approaches to zero in
peripheral and central collisions. As far as the charmonium concern, those that produced
from the initial collisions prior to the formation of the hot medium will not
present any significant elliptic flow, see the dot-dashed line in the upper panel of Fig.\ref{fig10}.
The regenerated $J/\psi$s, on the other hand, may inherit flow from the thermalized
charm quarks, as shown by the dashed line. Through interactions,
the initial anisotropy in the configuration space is converted into the
anisotropy in momentum space for all final hadrons. The full $J/\psi$ $v_{2}$ is shown as the solid line in
the upper panel of Fig.\ref{fig10}. Note that a sizable $v_{2}$ has been developed
at LHC. Unlike the light charged hadrons where $v_{2}$ persist with large values up to
high $p_{T}$ region, the $J/\psi$ $v_{2}$ quickly drops to the value from initial production
with $v_{2}\leq 1\%$.

Up to now, our discussion on $J/\psi$ $v_2$ is focused on the prompt $J/\psi$. Once we include the
$J/\psi$ from bottom quark decay, new scenarios arise. The lower panel of Fig.\ref{fig10} shows
the elliptic flow $v_{2}$ for inclusive $J/\psi$s at forward rapidity where the bottom decay is
included. As one can see in the plot, the $J/\psi$ $v_2$ from the decay of
the initially (or pQCD) produced bottom quarks quickly falls to close to zero at $p_T\sim 5$ GeV/c.
However, the $J/\psi$s from thermalized bottom decay
show significant value of $v_{2}$ at high $p_T$ region. At this
point, the error of the experimental data is too large to draw any conclusion of bottom quark thermalization.
Our calculation predicts that future precise $J/\psi$ $v_{2}$ at high $p_{T}$ will provide
important information on bottom production in high energy nuclear collisions.

\section{Conclusion}
We have studied the cold and hot nuclear matter effects on charmonium production
in relativistic heavy ion collisions. In the framework of a transport approach
for the charmonium motion plus a hydrodynamic description for the medium evolution,
we calculated the charmonium transverse momentum distribution and the corresponding ratios $R_{AA}$ and $r_{AA}$ in different
rapidity and centrality bins. Our model calculations agree reasonably well with the experimental
data at SPS, RHIC and LHC energies. We found that, in comparison with the often used nuclear modification factor
$R_{AA}=N_{AA}/\left(N_{coll}N_{pp}\right)$, based on the charmonium yield, the newly defined
nuclear modification factor
$r_{AA}=\langle p_T^2\rangle_{AA}/\langle p_T^2\rangle_{pp}$, based on the charmonium transverse
momentum distribution, is more sensitive to the nature of the hot medium and to the degree
of heavy quark thermalization. When the colliding energy increases from SPS to LHC, the
$p_T$ $r_{AA}$ changes dramatically from above unity and increasing as a function of
collision centrality to below unity and decreasing versus centrality. Different from the yield $R_{AA}$ which is strongly affected by both the cold and hot nuclear matter effects, the $p_T$ $r_{AA}$ is weekly affected by the shadowing effect. In addition, we observed
that the $J/\psi$ elliptic flow $v_{2}$ for minimum bias events evolves from almost zero at RHIC to a
large value similar to that of light hadrons at LHC. The root for
these observed dramatic changes is the formation of the strongly coupled quark-gluon plasma and the thermalization of charm quarks in high energy nuclear
collisions at the LHC energy.

\noindent {\bf Acknowledgement:} The work is supported by the NSFC, the MOST and the DOE grant Nos. 11221504, 11335005, 11275103, 2013CB922000, 2014CB845400, and DE-AC03-76SF00098.

\end{document}